\pdfoutput=1

\documentclass[11pt]{article}

\usepackage[preprint]{ACL}

\usepackage{times}
\usepackage{latexsym}
\usepackage{listings}
\usepackage{makecell}

\usepackage{enumitem}

\definecolor{codegreen}{rgb}{0,0.6,0}
\definecolor{codegray}{rgb}{0.5,0.5,0.5}
\definecolor{codepurple}{rgb}{0.58,0,0.82}
\definecolor{backcolour}{rgb}{0.95,0.95,0.92}
\definecolor{myxcodecolour}{rgb}{0.682,0.059,0.588}

\usepackage{listings}
\lstdefinelanguage{SQL}{
    keywords={CREATE, QUERY, FROM, SELECT, WHERE, ACCUM, PRINT, SUM, VERTEX, ALTER, RETURN, ATTRIBUTE, SPACE, PRIMARY, KEY, LOAD, VALUES, ORDER, GROUP, IS, NOT, NULL, THEN, BY, LIMIT, ADD, UPDATE, *, AND, OR, SET, REPLACE, FUNCTION, RETURNS, TRIGGER, BEGIN, END, IF, AS},
    keywordstyle=\color{codepurple}\bfseries,
    morecomment=[l]{--},
    morecomment=[l]{//},
    morecomment=[s]{/*}{*/},
    commentstyle=\color{codegreen}\itshape,
    morestring=[b]',
    sensitive=true,
    stringstyle=\color{red},
}

\usepackage[T1]{fontenc}

\usepackage[utf8]{inputenc}

\usepackage{microtype}

\usepackage{inconsolata}

\usepackage{graphicx}
\usepackage{multirow}
\usepackage{natbib} 
\usepackage[T1]{fontenc}
\usepackage{graphicx}
\usepackage{booktabs}
\usepackage{xspace}
\usepackage{amsmath} 
\usepackage{algorithm}
\usepackage{algorithmic}
\usepackage{tcolorbox}
\usepackage{cleveref}
\usepackage[dvipsnames]{xcolor}

\usepackage[misc]{ifsym}
\usepackage{pifont}
\newcommand{\cmark}{{\color{Green}\ding{51}}}  
\newcommand{\xmark}{{\color{Red}\ding{55} \xspace }}    

\newcommand{\name}{\textbf{\textit{Castle\xspace}}}
\newcommand{\SELECT}{\texttt{\color{Blue} SELECT\xspace}}

\title{\textit{Castle}: Causal Cascade Updates in Relational Databases with Large Language Models}
\author{
  \textbf{Yongye Su\textsuperscript{1}\thanks{Equal contributions.}}, 
  \textbf{Yucheng Zhang\textsuperscript{1}\footnotemark[1]}, 
  \textbf{Zeru Shi\textsuperscript{2}}, 
  \textbf{Bruno Ribeiro\textsuperscript{1}}, 
  \textbf{Elisa Bertino\textsuperscript{1}}
\\
\textsuperscript{1}Purdue University, USA \quad
\textsuperscript{2}Rutgers University, USA
\\
\small{
    \texttt{su311@purdue.edu, zhan4332@purdue.edu, zeru.shi@rutgers.edu, ribeirob@purdue.edu, bertino@purdue.edu}
    }
}

\begin{document}
\maketitle
\begin{abstract}

This work introduces \name, the first framework for schema-only cascade update generation using large language models (LLMs). Despite recent advances in LLMs for Text2SQL code generation, existing approaches focus primarily on \texttt{\color{Blue} SELECT} queries, neglecting the challenges of SQL update operations and their ripple effects. Traditional \texttt{\color{Blue} CASCADE UPDATE} constraints are static and unsuitable for modern, denormalized databases, which demand dynamic, context-aware updates. \name{} enables natural language instructions to trigger multi-column, causally consistent SQL \texttt{\color{Blue} UPDATE} statements, 
without revealing table content to the model. By framing \texttt{\color{Blue} UPDATE} SQL generation as a divide-and-conquer task with LLMs’ reasoning capacity, \name {} can determine not only
which columns must be directly updated, 
but also how those updates propagate through the schema, causing cascading updates — all via nested queries and substructures that ensure data confidentiality. We evaluate it on real-world causal update scenarios, demonstrating its ability to produce accurate SQL updates, and thereby highlighting the reasoning ability of LLMs in automated DBMS.

\end{abstract}



\section{Introduction}

Relational Database Management Systems (RDBMS) are the foundation of modern information systems, providing reliable storage and efficient retrieval for critical business data. Natural language interfaces to databases, such as Text2SQL approaches, have enabled users to pose complex questions and retrieve answers via generated SQL queries~\citep{zhong2017seq2sql, xu2017sqlnet, yu-etal-2018-spider, guo2019towards, wang2019rat, scholak2021picard}. However, these efforts have primarily focused on generating retrieval-focused (\texttt{\color{Blue} SELECT}) queries.

To create a complete natural language interface, it is essential to also generate SQL \texttt{\color{Blue} UPDATE} commands. While accuracy challenges persist~\cite{yao2025taubench,dinsql_nips2023}, recent work by \citet{li-etal-2024-multisql} has pushed towards a more comprehensive Text2SQL framework, incorporating a broader set of SQL commands, including SQL \texttt{\color{Blue} UPDATE}s. Nevertheless, a significant challenge arises with cascade updates, where a change in one record requires automatic propagation of modifications to related records, causing a ``ripple effect'' in the database, particularly in high-performance denormalized databases~\cite{kimball2013data, balmin2005storing} (see \Cref{tab:denorm-vs-norm}). The advent of massively distributed systems and real-time analytics has increasingly led designers to adopt \textbf{denormalized} schemas~\cite{kimball2013data}, where relational dimensions are flattened to reduce expensive join operations and meet performance targets.
For instance, consider a denormalized database of soccer players: in a denormalized database, the table of players also have records about the clubs, such as the club's name and coach. After a player joins a new club (update club name), their coach name needs to be updated, but this information resides in the table of clubs. Thus, we need to update corresponding table entities in a cascading fashion. 

\begin{table*}[t]
\centering
\small
\begin{tabular}{@{}p{3cm}|p{6cm}p{5cm}@{}}
\toprule
\textbf{Scenario} & \textbf{Denormalized Schema} & \textbf{Normalized Schema} \\
\hline
{\color{Black}SELECT Query}& \cmark\ {\color{Black}No joins required} & \xmark\ {\color{Black}Requires multi-table joins} \\

{\color{Black}Update w/ Cascade} & \xmark\ {\color{Black} Hard to trace and inconsistent updates risks} & \cmark\ {\color{Black}Easy with relational foreign keys} \\
\bottomrule
\end{tabular}
\caption{Comparison between denormalized and normalized schemas in practice, symbols indicate relative advantages (\cmark) and disadvantages (\xmark).}
\label{tab:denorm-vs-norm}
\end{table*}

 This work addresses the task of improving Text2SQL \texttt{\color{Blue} UPDATE}s, with a focus on \textbf{cascade updates}. We create two cascade update benchmarks using public datasets to test the ability of Text2SQL methods to issue correct update commands under cascades over more than 1 million records. We then introduce \name, a new framework designed to enable large language models (LLMs) to generate SQL update commands that execute intended modifications and automatically handle causal-driven cascade updates securely and efficiently. A key challenge is uncovering real-world causal relationships between updated fields and other fields. Our approach also prioritizes preserving data confidentiality by utilizing nested query construction instead of table data augmented generation. Our experimental results demonstrate the effectiveness of our framework, achieving up to 85\% correct updates in our benchmark tasks, consistently outperforming the best baselines, which reach at most 80\% correctness, and are often much lower (down to 52\%) in complex scenarios.

 Our main contributions are as follows:
 \begin{itemize}[leftmargin=*]
    \item \name {} is the first framework tailored specifically for SQL cascade update operations. It treats SQL cascade updates as causal reasoning tasks. With nested structured subqueries, \name {} generates update commands without exposing raw table data, mitigating privacy risks inherent in current table-augmented approaches. 
    \item We propose two datasets for cascading updates that are 100\% based on causal relationships from the real world with more than 1 million records. 
    \item \name {} is the first work that systematically studies and evaluates the LLM-assisted SQL Trigger management and generation.
\end{itemize}

\section{\name}

\paragraph{Research Question} Modern database designs often exhibit performance-driven redundancies, which complicate update operations. Specifically, we are interested in the question: \textbf{Can Large Language Models generate accurate cascade update queries correctly given only the database schema?} This research question gives rise to two fundamental challenges:

\noindent
\textbf{Identifying Update Targets (C1)}. When generating \texttt{\color{Blue} UPDATE}  queries, it is essential to determine the specific columns that require modification. From the perspective of LLMs, this involves not only identifying the target column(s) specified in the natural language instruction but also recognizing potential related updates to other columns, which can vary on a case-by-case basis.


\noindent
\textbf{Determining Update Values (C2)}. After identifying the columns to update, the cascade update operation must determine the new values to assign to the corresponding columns. Since these values are not explicitly provided in the natural language instructions, they may need to be inferred from other data entities, posing a significant challenge for LLMs.


\paragraph{Motivating exampling using our Soccer Transfer task.} In order to illustrate the challenges, we first introduce our Soccer Transfer dataset schema in \Cref{schema-Soccer Transfer}. Consider the instruction: ``Lionel Messi has transferred from Barcelona (code: fc-barcelona) to Paris Saint-Germain (code: fc-paris-saint-germain), update his information.'' In response to this instruction, an LLM should not only update the columns directly mentioned (e.g., updating the club\_name from ``Barcelona'' to ``Paris Saint-Germain''), but also infer and update causally related columns. For instance, it should update the coach\_name from ``Ronald Koeman'' to ``Mauricio Pochettino'', reflecting the change in team affiliation. This example highlights the need for LLMs to capture complex causal relationships within the data schema to generate accurate and comprehensive updates.

\paragraph{Proposed Method.} To enable LLMs to perform causally-driven cascading updates from natural language instructions, without sending table content data to models that compromise data confidentiality, in this section, we introduce \name \xspace 
(see Figure~\ref{fig:pipeline}), a multi-stage workflow addressing the aforementioned challenge via divide-and-conquer chain-of-thoughts (DC-CoT) with zero-shot samples. \name \xspace orchestrates the generation and execution of SQL \texttt{\color{Blue} UPDATE} queries from natural language instructions, maintaining data consistency via causal reasoning and robust query construction. The entire process is detailed in Algorithm~\ref{alg:castle} and described in what follows.

\begin{algorithm}[h]
\caption{\textsc{Workflow of C.A.S.T.L.E.} }
\label{alg:castle}
\begin{algorithmic}[1]
\REQUIRE Natural language update instruction $x$, table schema $S$

\STATE \textbf{C. Column Identification:} \label{line:Column_Identification}
\STATE \quad Based on $S$, extract directly mentioned target column(s) $C_{direct}$ and table from $x$

\STATE \textbf{A. Attribute Dependency Analysis:} \label{line:ada}
\STATE \quad Use schema $S$ and reasoning over $C_\text{direct}$ to infer causally dependent columns $C_\text{cascade}$

\STATE \textbf{S. Subquery Planning:}  \label{line:Subquery}
\FOR{each $c \in C_\text{cascade}$}
    \STATE Generate subquery $q_c$ to retrieve correct value for $c$ based on $S$
\ENDFOR

\STATE \textbf{T. Trigger Maintenance:}\label{line:Trigger}
\FOR{each derived aggregate column $c \in C_\text{cascade} \cup C_\text{direct}$}
    \STATE Check trigger for maintaining derived $c$  
    \IF{trigger is missing}
        \STATE Generate SQL trigger $t_c$ (via schema-based causal reasoning)
        \STATE Deploy trigger $t_c$ into the database to maintain real-time consistency
    \ENDIF
\ENDFOR

\STATE \textbf{L. Logical Query Composition:} \label{line:lqc}
\STATE \quad Combine $C_\text{direct}$ values from x and valid subquery $q_c$ into final SQL \texttt{UPDATE} query $q$

\STATE \textbf{E. Execution:} \label{line:exec}
\STATE \quad Execute \texttt{UPDATE} query $q$ on target data table.

\end{algorithmic}
\end{algorithm}

\subsection{Skeleton: Identifying Columns for Update}
\begin{figure*}[t]
    \centering
    \includegraphics[width=1\linewidth]{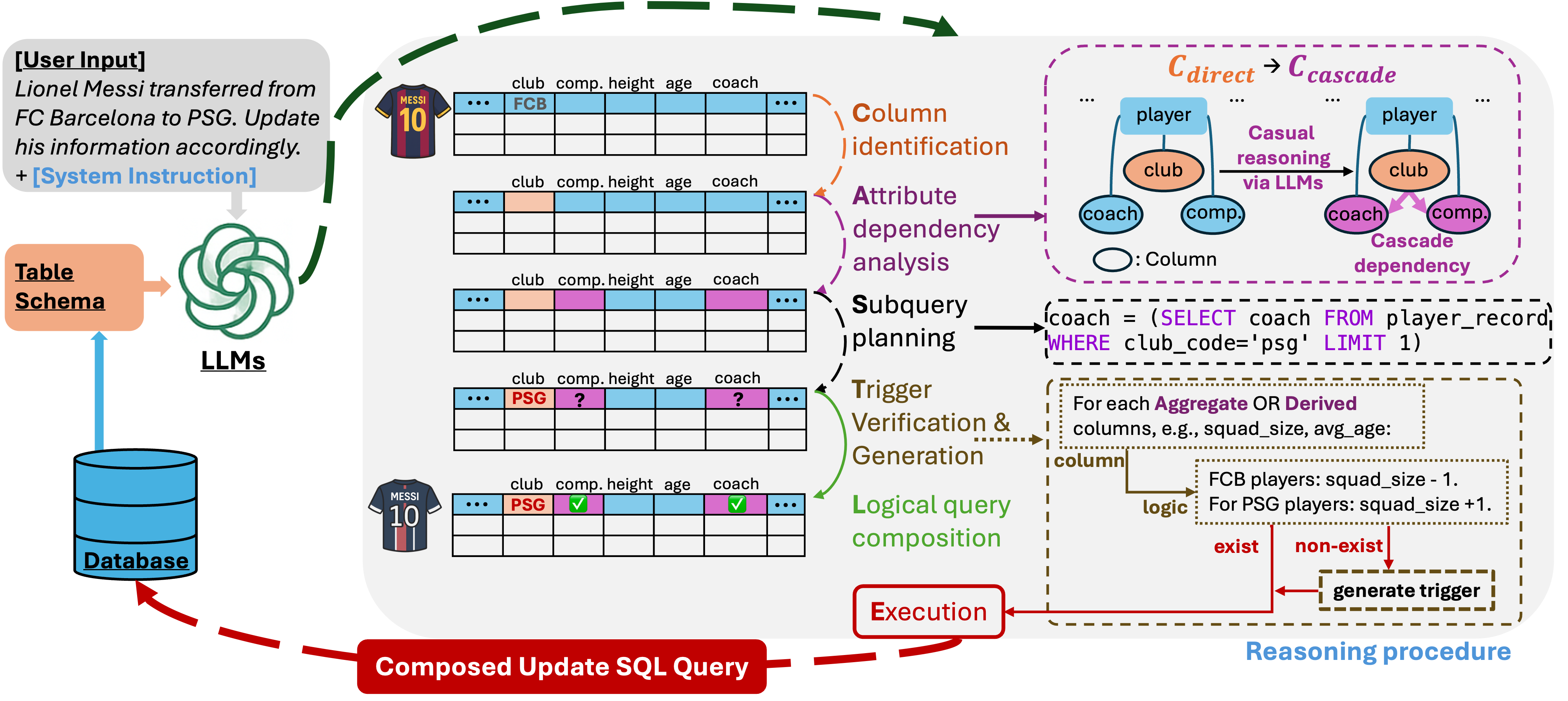}
    \caption{A sample workflow of \name \xspace illustrates how a single transaction with natural language instructions as input is completed as a cascading update SQL transaction, incorporating the reasoning procedures of LLMs alongside the database.}
    \label{fig:pipeline}
\end{figure*}

As the first critical step to update the data in tables, \name \xspace needs to accurately identify 
the columns to be updated from a natural language instruction. The expected result of this step is the skeleton of SQL \texttt{\color{Blue} UPDATE} queries. \name \xspace distinguishes between two types of columns from the database table schema:


\paragraph{Columns to be directly updated ($C_\text{direct}$).} Using the given table schema and natural language instructions from general users, \name \xspace identifies explicitly mentioned columns that need to be updated with data provided by the users. For example, like shown in Figure~\ref{fig:pipeline}, given the instruction ``Lionel Messi has transferred club from Barcelona (code: fc-barcelona) to Paris Saint-Germain (code: fc-paris-saint-germain)'' the directly related columns \texttt{club\_name} and \texttt{club\_code} are explicitly identified by \name. This procedure corresponds to Line~\ref{line:Column_Identification} in Algorithm~\ref{alg:castle}.

\paragraph{Causally-dependent columns to be cascade updated ($C_\text{cascade}$).} \name \xspace applies the causal reasoning capabilities of LLMs via structured instruction to identify implicitly affected columns in each transaction (e.g., \texttt{\color{Blue} UPDATE}) in the database. For instance, updating a player's club may also require updating dependent columns such as this player's \texttt{coach} or \texttt{competition}, as depicted in Figure~\ref{fig:pipeline}. This procedure corresponds to Line~\ref{line:ada} in Algorithm~\ref{alg:castle}.

After the identification of the columns to be directly updated and/or cascade updated,
a skeleton of the SQL \texttt{\color{Blue} UPDATE} query is ready as shown in Code~\ref{lst:update-skeleton}.
Now the question remains to be ``what to update''.

\begin{lstlisting}[language=SQL, caption={An example of generated SQL \texttt{\color{Blue} UPDATE} skeleton for table \texttt{player\_record}. Question marks here serve as placeholders for later subqueries or trigger maintenance. }, label={lst:update-skeleton}]

UPDATE player_record
SET
    "club_name" = 'Paris Saint-Germain', -- directly update
    "club_code" = 'psg', -- directly update
    ...
    "stadium_name" = ?, -- causally-dependent column 
    "competition_country" = ?, -- causally-dependent column
    ...
    "foreigners_percentage" = ?, -- aggregate and derived column
    "squad_size" = ? -- aggregate and derived column
    ...
WHERE
    "player_code" = 'lionel-messi';



\end{lstlisting}

\subsection{Subquery Planning: Evidence-grounded Updates}
After identifying the columns to update, \name \xspace proceeds to handle causally-dependent updates securely through structured subquery planning. Instead of exposing actual table content data to LLMs, \name {} only provides the table schema along with system instructions as input to LLMs integrated with the system. For each causally dependent column ($c \in C_\text{cascade}$) that needs to be updated, the LLM generates \texttt{\color{Blue} SELECT} subqueries to fetch the correct value to update. Each subquery expects a result cardinality of at most one, guaranteeing legal and precise 
updates without data exposure, shown as an example in the dashed box in the middle of the right column of Figure~\ref{fig:pipeline}.

Moreover, prior to query composition and execution, each generated subquery undergoes syntax checking to ensure its cardinality and compatibility with database-specific dialect rules, explicitly validating clauses like \texttt{\color{Blue} LIMIT} (e.g., PostgreSQL and MySQL) or \texttt{\color{Blue} TOP} (e.g., Microsoft SQL Server), thereby minimizing syntax-related runtime errors.

\subsection{Trigger Verification and Generation}
Aggregate and derived columns (e.g., counts, averages, or percentage metrics) or materialized tables in relational databases often reflect precomputed summarized information crucial for fast analytical retrieval~\cite{jugel2016vdda} in Business Intelligence (BI). When underlying data changes occur (e.g., player transfers or retail records change), traditional ETL-based methods may rely on scheduled batch data recomputing to maintain data consistency. \name \xspace addresses this through automatic verification and generation of SQL \texttt{\color{Blue} TRIGGER}, which efficiently maintains these derived metrics in real-time upon data modification events. This mechanism provides robustness, consistency, and efficiency for event-driven causal cascade updates.

\paragraph{Trigger Verification.}
Given columns identified either directly or via causal reasoning ($C_\text{cascade} \cup C_\text{direct}$), \name \xspace first checks existing SQL Triggers from the database. Checking the existence and syntax of triggers based on the current schema from the system's maintained triggers table/view.

\paragraph{Trigger Generation.}
If the verification reveals missing triggers, \name \xspace utilizes the schema-based causal reasoning capability of LLMs to dynamically generate efficient SQL trigger scripts with functions to maintain data consistency from events. The triggers are designed explicitly to accurately reflect updated derived (or aggregate) metrics like \texttt{squad\_size}, \texttt{average\_age},  and \texttt{foreigners\_percentage}. Such one-time effort can automatically and consistently propagate changes without further manual intervention, thereby enhancing data integrity across the database.

In \name, generated triggers are subsequently deployed to the database and seamlessly integrated into transaction workflows. This proactive trigger management significantly reduces runtime computational overhead and LLM token usages, but also guarantees consistency for aggregate and derived data metrics in databases.

\subsection{Compatibility with other systems}

\name{} is designed to be broadly compatible with industry standard DBMS such as PostgreSQL~\footnote{https://www.postgresql.org/}, MySQL~\footnote{https://www.mysql.com/}, Microsoft SQL Server~\footnote{https://www.microsoft.com/en-us/sql-server}, and so on. On the one hand, it runs purely at the schema level, without requiring system-specific modules, extensions, and configurations (e.g., indices, paging, caching). Thus, it can be seamlessly integrated into existing data infrastructure without requiring modifications or additional extensions to the underlying database engine.

\section{Experiments}
We run our experiment on PostgreSQL 17 hosted on Neon~\footnote{https://www.neon.tech/}, a serverless database platform built on AWS Aurora Postgres, with different LLMs integrated in the system workflow, including ChatGPT-4o~\footnote{https://openai.com/index/hello-gpt-4o/}, LLaMA-3.1-8B-Instruct~\footnote{https://huggingface.co/meta-llama/Llama-3.1-8B-Instruct} and Qwen2.5-7B-Instruct~\footnote{https://huggingface.co/Qwen/Qwen2.5-7B-Instruct} as SQL generators.

\subsection{Dataset}
To evaluate our approach to causally driven cascade updates in structured relational databases, we utilize two real-world datasets:

\begin{table}[h]
\centering
\small
\begin{tabular}{@{}lccc@{}}
\toprule
\textbf{Dataset} & \textbf{Size} & \textbf{\#Columns} & \textbf{Data Type} \\
\midrule
Soccer Transfer     & 1M+     & 28                  & Date, Text, Numeric   \\
UCI Retail & 541K       & 12                  & Date, Text, Numeric  \\
\bottomrule
\end{tabular}
\caption{Overview of relational datasets used in our experiment evaluation.}
\label{tab:dataset-overview}
\end{table}

\paragraph{Soccer Transfer Dataset.} A complete and comprehensive dataset recording over a million football player appearances and transfers worldwide yearly, including personal details, club and national team affiliations, transfer histories, and performance statistics. The relational structure allows users to model complex dependencies. Our ground truth of one year's update information is based on the following year's player record. We compare every pair of adjacent year records and find out those players who changed their club, such difference provides the club update information for us to extract the fact from the later year and evaluate LLM's ability to perform causal cascade update on the earlier year's record~\footnote{https://www.kaggle.com/datasets/davidcariboo/player-scores}. 

\paragraph{UCI Online Retail II Dataset.}  This dataset contains over half a million transactional records from a UK-based online retailer to worldwide customers, covering sales and returns over two years. Each record includes attributes like invoice number, product code, quantity, invoice date, unit price, customer ID, and country~\cite{online_retail_ii_502}. We augment this dataset by: (1) Deriving the \textbf{Quarter} from the invoice date, corresponding to the quarter report in a materialized view (table) for faster data retrieval. (2) Mapping \textbf{Country} to \textbf{Region} for geographical analysis.

\subsection{Evaluation Metrics}
In retrieval-focused Text2SQL, one of the metrics is Execution Accuracy, which measures how close the results of the generated SQL query are to the ground truth results. In MultiSQL~\cite{li-etal-2024-multisql}, the evaluation metric for update operations in databases is state comparison, which directly compares the whole content of two database states (before and after update operation), returning a binary result of 0 (different) or 1 (same). Unlike the metric in MultiSQL, according to our workflow design, while performing database update operations, the first question is to identify the data to update.

Thus, for the update operations, given the causal and ripple-effect nature of cascade updates, we evaluate the model's reasoning capability of the causal cascade update scenario through \textbf{recall} (how many of the truly needed updates were found), with breakdown into those directly updated columns and cascade updated columns.  This metric assesses the model's ability to holistically reason about multiple column dependencies within and outside the data table, quantifying how many of the truly needed updates were identified after the LLM reasoning procedure, and corresponds to evaluating the proposed "where to update" challenge in this work.

\begin{equation}
\text{Recall} = \frac{|\delta_{\text{Identified \ Cell \ to\ Update }}|} {|\Delta_\text{Total \ Cell \ Requiring \ Updates}|} 
\end{equation}



On the other hand, after we quantify the columns identified by LLMs to update, we also need to know the proportion of correctly updated causal columns among those targeted for update by the model. For example, after identifying columns to update, if one model fails to update all required columns, it has two kinds of errors: Type I (unnecessary update) errors and Type II (missed update) errors, either could happen when conducting updates to the database. The \textbf{F1-score} summarizes both aspects, indicating the model's overall effectiveness in both identifying and accurately updating causally dependent columns.

\begin{table*}[h]
\centering
\renewcommand{\arraystretch}{1.2}
\setlength{\tabcolsep}{10pt}
\begin{tabular}{l|ccc}
\toprule
\textbf{Method $\backslash$ Model} & \textbf{ChatGPT-4o} & \textbf{LLaMA-3.1-8B-Instruct} & \textbf{Qwen2.5-7B-Instruct} \\
\midrule
\textbf{\name (w/o data)} &
\makecell{ \textbf{99.96} $\mid$ \textbf{89.39} $\mid$ \textbf{80.82}} &
\makecell{ \textbf{99.49} $\mid$ \textbf{88.62} $\mid$ \textbf{79.45} }  &
\makecell{ \textbf{96.32} $\mid$ \textbf{87.58} $\mid$ \textbf{77.45} } \\

\textbf{Multi-SQL (w/ data)} &
\makecell{ 99.25 $\mid$ 88.62 $\mid$ 79.45 } &
\makecell{ 95.50 $\mid$ 82.93 $\mid$ 70.76 } &
\makecell{ 90.41 $\mid$ 84.01 $\mid$ 72.22 } \\

\textbf{Baseline (w/ data)} &
\makecell{ 99.18 $\mid$ 88.35 $\mid$ 79.00 } &
\makecell{ 99.32 $\mid$ 88.19 $\mid$ 78.76 } &
\makecell{ 96.10 $\mid$ 83.41 $\mid$ 71.98} \\
\bottomrule
\end{tabular}

\begin{tabular}[h]{l|ccc}
\toprule
\textbf{Method $\backslash$ Model} & \textbf{ChatGPT-4o} & \textbf{LLaMA-3.1-8B-Instruct} & \textbf{Qwen2.5-7B-Instruct} \\
\midrule
\textbf{\name (w/o data)} &
\makecell{ \textbf{95.93} $\mid$ \textbf{86.45} $\mid$ \textbf{83.55}} &
\makecell{ \textbf{91.25} $\mid$ \textbf{84.32} $\mid$ \textbf{81.63} }  &
\makecell{ \textbf{89.23} $\mid$ \textbf{84.03} $\mid$ \textbf{81.90} } \\

\textbf{Multi-SQL (w/ data)} &
\makecell{ 93.68 $\mid$ 85.83 $\mid$ 81.68 } &
\makecell{ 79.87 $\mid$ 71.83 $\mid$ 73.39 } &
\makecell{ 81.34 $\mid$ 73.47 $\mid$ 75.90 } \\

\textbf{Baseline (w/ data)} &
\makecell{ 92.15 $\mid$ 84.76 $\mid$ 80.58 } &
\makecell{ 75.78 $\mid$ 69.12 $\mid$ 65.13} &
\makecell{ 74.89 $\mid$ 72.09  $\mid$ 64.63} \\
\bottomrule
\end{tabular}

\caption{Evaluation of update performance across models with methods on \textbf{Soccer Transfer} and \textbf{Retail} dataset, respectively. Each cell reports: Recall, F1-score, and cell-wise correct rate of \textbf{directly-updated columns.}}
\label{tab:direct-update-results}
\end{table*}

\begin{table*}[h]
\centering
\renewcommand{\arraystretch}{1.2}
\setlength{\tabcolsep}{10pt}
\begin{tabular}{l|ccc}
\toprule
\textbf{Method $\backslash$ Model} & \textbf{ChatGPT-4o} & \textbf{LLaMA-3.1-8B-Instruct} & \textbf{Qwen2.5-7B-Instruct} \\
\midrule
\textbf{\name (w/o data)} & 
\makecell{ \textbf{99.95}  $\mid$ \textbf{85.25} $\mid$ \textbf{85.21}} &
\makecell{ \textbf{80.84} $\mid$ \textbf{81.24} $\mid$ \textbf{80.43} } &
\makecell{ \textbf{75.52} $\mid$ \textbf{72.01} $\mid$ \textbf{67.93} } \\

\textbf{Multi-SQL (w/ data)} &
\makecell{ 52.21 $\mid$ 68.58 $\mid$ 52.16 } &
\makecell{ 50.31 $\mid$ 68.97 $\mid$ 52.17} &
\makecell{ 52.39 $\mid$ 68.73 $\mid$ 55.39 } \\

\textbf{Baseline (w/ data)} &
\makecell{ 52.16 $\mid$ 68.53 $\mid$ 52.09 } &
\makecell{ 50.09 $\mid$ 67.55 $\mid$ 52.00 } &
\makecell{ 51.12 $\mid$ 65.80 $\mid$ 52.06 } \\
\bottomrule
\end{tabular}

\begin{tabular}{l|ccc}
\toprule
\textbf{Method $\backslash$ Model} & \textbf{ChatGPT-4o} & \textbf{LLaMA-3.1-8B-Instruct} & \textbf{Qwen2.5-7B-Instruct} \\
\midrule
\textbf{\name (w/o data)} &
\makecell{ \textbf{93.03} $\mid$ \textbf{90.02} $\mid$ \textbf{85.93}} &
\makecell{ \textbf{89.32} $\mid$ \textbf{81.36} $\mid$ \textbf{83.31} }  &
\makecell{ \textbf{87.98} $\mid$ \textbf{83.21 } $\mid$ \textbf{84.02} } \\

\textbf{Multi-SQL (w/ data)} &
\makecell{  89.10 $\mid$ 74.12 $\mid$ 69.58 } &
\makecell{  83.80 $\mid$ 58.80 $\mid$ 56.09 } &
\makecell{  85.91 $\mid$ 65.01 $\mid$ 61.23 } \\

\textbf{Baseline (w/ data)} &
\makecell{ 88.45 $\mid$ 76.16 $\mid$ 70.15  } &
\makecell{ 82.42 $\mid$ 55.14 $\mid$ 53.76 } &
\makecell{ 85.04 $\mid$ 66.10 $\mid$ 70.49 } \\
\bottomrule
\end{tabular}
\caption{Evaluation of update performance across models with methods on \textbf{Soccer Transfer} and \textbf{Retail} dataset, respectively. Each cell reports: Recall, F1-score, and cell-wise correct rate of \textbf{causal cascade updated columns} (without derived values).}
\label{tab:causal-update-results}
\end{table*}




Besides, correctness is the final requirement in our task settings, where correct results would justify the usability of our proposed workflow. Thus, we introduce cell-wise correctness (CC), which evaluates fine-grained correct rate on how accurately the model updates each individual cell (i.e., each column within each row) across the entire database after applying one natural language update instruction. The Cell-wise correctness (CC) is defined as follows:

\begin{equation}
\text{CC} = \frac{|\delta_{\text{Correct \ Cell \ Updated }}|} {|\Delta_\text{Total \ Cell \ Requiring \ Updates}|} 
\end{equation}

Last but not least, in order to further study how causal dependent columns are correctly updated aside from direct updates, we further break these metrics down into two complementary components: columns explicitly mentioned in the natural language instruction ($C_\text{direct}$) and cascading columns inferred from causal or structural dependencies ($C_\text{cascade}$).


\subsection{Results}

\begin{table*}[h]
\centering
\begin{tabular}{lccc}
\toprule
\textbf{Method / Model} & \textbf{ChatGPT-4o} & \textbf{LLaMA-3.1-8B-Instruct} & \textbf{Qwen2.5-7B-Instruct} \\
\name {} (Trigger only)    & 83.31 & 79.37 & 77.29 \\
\textbf{\name {} w/ trigger}       & 83.84 & 79.91 & 73.53 \\
\bottomrule
\end{tabular}

\begin{tabular}{lccc}
\toprule
\textbf{Method / Model} & \textbf{ChatGPT-4o} & \textbf{LLaMA-3.1-8B-Instruct} & \textbf{Qwen2.5-7B-Instruct} \\
\midrule
\name {} (Trigger only)    & 87.64 & 77.90 & 81.01 \\

\textbf{\name {} w/ trigger}       & 86.81 & 79.51 & 81.78 \\
\bottomrule
\end{tabular}
\caption{Evaluation of LLM-generated \texttt{TRIGGER} via cell-wise correctness, bottom row represents \name's {} average cell-wise correctness over all data columns and tables with integrated trigger generation mechanism.}
\label{tab:trigger-correctness}
\end{table*}

While \name {} does not provide a \texttt{\color{Blue} SELECT}-like query result as output, we evaluated our update results by querying them and comparing them with the corresponding ground truth, as soon as the update operation occurred in the database. In Table~\ref{tab:direct-update-results} and Table~\ref{tab:causal-update-results}, we present the evaluation results of update performance for directly-updated columns ($C_{\text{direct}}$) and causal cascade-updated columns ($C_{\text{cascade}}$, but without derived columns for \texttt{\color{Blue} TRIGGER}s to maintain) across three representative LLMs and three SQL generation methods: \name {} (ours, schema-only), MultiSQL (content-augmented), and baseline method (content-augmented). 

In addition to direct update commands, we also evaluated the ability of LLMs to generate correct SQL \texttt{\color{Blue} TRIGGER} statements within our workflow that enforce ripple-effect data consistency with our causal cascade \texttt{\color{Blue} UPDATE} queries. The generated trigger is considered correct if, once after deployment, it consistently maintains the correctness of the summary table right after transactional updates, as compared with ground truth using cell-wise correctness. The result is shown in Table~\ref{tab:trigger-correctness} alongside \name {} having trigger generated in the system, both of their cell-wise correctness are the average rate of experiments conducted 100 times.  Our experiments demonstrate that data consistency can be maintained automatically and robustly, even across complex, multi-row updates.

\subsection{Discussion}


In our evaluation, we measured "where to update" via the recall metric of cascade reasoning, and also "what to update" with F1-score and cell-wise correctness metrics.

The recall metric in our experiment directly evaluates the LLM's ability to correctly identify where to update in the given table schema(s), i.e., which columns (both direct and causal/cascade) should be updated based on natural language instructions. Our schema-only approach, \name, consistently achieves the highest recall across all models, particularly with GPT-4o, outperforming content-augmented baseline methods on both directly and causally updated columns. This demonstrates \name’s ability to reason over schema semantics and relationships without access to table content.

F1-score and cell-wise correctness indicate the model’s proficiency in determining what values to update and producing the correct SQL subqueries filling the outer skeleton. We discover \name {} performs consistently better, especially in scenarios with larger or more complex schemas, or having column interdependencies (as in the Retail dataset), where achieving high cell-wise correctness becomes more challenging.

Table~\ref{tab:trigger-correctness} presents the evaluation of LLM-generated triggers for maintaining aggregate or materialized columns in real-time. The results show that triggers generated by \name {} (both standalone and integrated) are comparable to LLM-generated complex SQL queries. However, triggers are instantly activated and only require one-time effort that can provide long-term, automatic consistency without requiring repetitive code generation or human intervention. This finding shows the potential of integrating LLM-based trigger generation into modern DBMS.

\section{Related Work}

\subsection{Text2SQL}
In general, Text2SQL (or NL2SQL) takes a given natural language text query as a task, 
generates the task-specific SQL queries~\cite{DBLP:conf/edbt/MitsopoulouK25, liu2024survey, ma2025sql}, and compares the query result table with the groundtruth provided by baselines such as Spider~\cite{lei2024spider}, BIRD~\cite{li2023llm}, and CoSQL~\cite{yu-etal-2019-cosql}. However, until recently, Text2SQL datasets contained almost exclusively \SELECT \xspace queries, and update operations have been little investigated in Text2SQL research. The recent MultiSQL approach~\cite{li-etal-2024-multisql} supports
generating simple direct update commands with table content provided to LLMs for SQL generation~\cite{shen2024select, he2025star}. Moreover, the authors of MultiSQL provided a benchmark dataset that includes \texttt{\color{Blue} UPDATE} commands. However, those LLM-generated virtual data lack quality and real-world verifiability for causal relationships. In our work, two real verifiable datasets from different domains with different structures are used to verify our proposed method in causal cascade update.



In Text2SQL tasks, another common shortcoming for accurate query generation is the need for table context to understand natural language intents better~\cite{sun-etal-2018-semantic}. If table content from the actual query table is provided, it could significantly increase the accuracy of the generated query ~\cite{DBLP:conf/edbt/MitsopoulouK25}. However, such an approach undermines data confidentiality. By contrast, our approach achieves schema-only reasoning without having to send table content to LLMs. 

Recent approaches, such as CHASE-SQL~\cite{pourreza2025chasesql}, focus on improving SQL query generation via divide-and-conquer strategies and chain-of-thought (CoT)~\cite{NEURIPS2022_9d560961}. Our approach also applies this reasoning strategy to address the challenges mentioned in the paper's introduction section. 

\subsection{LLM-Assisted Data Wrangling}
Recent research has expanded the role of LLMs from purely generating SQL statements to performing broader data wrangling tasks. For instance, CodexDB~\cite{trummer2022codexdb} leverages Codex models to automate database interactions, demonstrating LLM capabilities for diverse database operations. TableLLM~\cite{zhang2025tablellmenablingtabulardata} is a dedicated model for document-level (lightweighted) spreadsheet manipulations, including insert, update, and delete operations. However, these operations require the whole table to be fed to the context window of LLMs. In addition, each operation is generated 
in isolation, 
neglecting cascades or multi-record dependencies. 

Unlike these methods, \name \xspace combines LLM-generated SQL code with a schema-driven reasoning process, systematically managing those causal cascade updates in denormalized schemas.

\subsection{Ripple Effects in Knowledge Editing}
Knowledge editing~\cite{mitchell2021fast, meng2022locating, meng2022mass} in LLMs aims to update specific factual information within a model without necessitating retraining. However, such interventions often lead to ``ripple effects''~\cite{cohen2024evaluating}, where modifications to one fact inadvertently influence related or unrelated knowledge within the model. \cite{cohen2024evaluating} introduced the RippleEdits benchmark to assess these effects, revealing that current editing methods frequently fail to ensure consistent knowledge updates, thereby compromising the model's reliability. Further analysis in GradSim~\cite{qin2024does} identified gradient similarity (GradSim) as a key indicator of ripple effects, demonstrating a strong positive correlation between GradSim and the successful propagation of edits. To address these challenges, \cite{zhao2024ripplecot} proposed RippleCOT, an in-context learning approach that integrates chain-of-thought reasoning to enhance the accurate dissemination of edits across related facts. Collectively, these works underscore the complexities inherent in knowledge editing for LLMs and highlight the necessity for advanced methods to manage unintended ripple effects.

While prior research has primarily examined ripple effects within LLMs~\cite{cohen2024evaluating, qin2024does}, our work shifts focus to the ripple effects occurring in external databases that serve as knowledge bases for LLMs. In retrieval-augmented generation (RAG) pipelines, effectively managing ripple effects during data retrieval by the DBMS can significantly enhance the accuracy and reliability of downstream LLM outputs~\cite{shi2024retrieval, zhao2024ripplecot}. Our approach offers a complementary perspective to existing model-level interventions, emphasizing the importance of database-level strategies in mitigating unintended ripple effects.

\section{Conclusion}

\name{} addresses the challenge in causal-driven cascade updates with respect to both ``where to update'' and ``what to update''. It also demonstrates that general pre-trained LLMs can reason over schema structures to perform cascade-consistent SQL updates without requiring access to table contents, thus 
providing 
a broader, trustworthy, structured LLM reasoning for general data systems and code generation.

\section*{Limitations}
Like most practical analytical queries and business intelligence (BI) workloads, we also assume the scenarios where data and its derived values are stored in a single unified database for optimized query performance. We thus do not consider federated or multi-database environments. Additionally, we do not consider multi-hop post-cascade updates due to the absence of real-world, verifiable datasets that reliably capture such propagation chains. Lastly, our work considers natural language instructions for the database to be explicit, which are generated via a unified script as part of input to LLMs; our study does not consider instructions in different languages aside from English.

\bibliography{custom}

\newpage
\appendix

\section{Table Schema}
\begin{tcolorbox}[title={Soccer Transfer Database Schema}, label={schema-Soccer Transfer}]
\scriptsize
\begin{lstlisting}
CREATE TABLE IF NOT EXISTS {table_name} (
    player_id                 SERIAL PRIMARY KEY,
    player_code               VARCHAR(100),
    first_name                VARCHAR(100),
    last_name                 VARCHAR(100),
    full_name                 VARCHAR(255),
    date_of_birth             VARCHAR(100),
    age                       NUMERIC,
    height                    DECIMAL(4,2),
    citizenship               VARCHAR(100),
    position                  VARCHAR(100),
    foot                      VARCHAR(20),
    contract_expires          VARCHAR(100),
    social_media              JSONB,
    birthplace_city           VARCHAR(100),
    birthplace_country        VARCHAR(100),
    club_code                 VARCHAR(100),
    club_name                 VARCHAR(255),
    squad_size                NUMERIC,
    average_age               DECIMAL(4,2),
    foreigners_number         NUMERIC,
    foreigners_percentage     DECIMAL(5,2),
    national_team_players     NUMERIC,
    stadium_name              VARCHAR(255),
    stadium_seats             VARCHAR(50),
    net_transfer_record       VARCHAR(50),
    coach_name                VARCHAR(255),
    competition_code          VARCHAR(50),
    competition_type          VARCHAR(50),
    competition_country       VARCHAR(100),
    competition_seasoned_href TEXT
    );
\end{lstlisting}
\end{tcolorbox}

\begin{tcolorbox}[title={UCI Retail Database Schema}]
\scriptsize
\begin{lstlisting}
CREATE TABLE IF NOT EXISTS {table_name} (
        stockcode TEXT,
        description TEXT,
        quantity INTEGER,
        country TEXT,
        region TEXT,
        y2010q4_quantity INTEGER,
        y2011q1_quantity INTEGER,
        y2011q2_quantity INTEGER,
        y2011q3_quantity INTEGER,
        y2011q4_quantity INTEGER,
        PRIMARY KEY(stockcode, country)
    );
\end{lstlisting}
\end{tcolorbox}



\section{Trigger}

\begin{lstlisting}[language=SQL, caption={An example query of listing Triggers names and corresponding tables in a PostgreSQL database.}, label={lst:checking_trigger}]
SELECT event_object_table AS table_name, trigger_name
FROM information_schema.triggers
GROUP BY table_name, trigger_name
ORDER BY table_name, trigger_name;
\end{lstlisting}

\begin{lstlisting}[language=SQL, caption={An example query of checking Trigger on table \texttt{player\_record}}, label={lst:checking_trigger}]
SELECT tgname
FROM pg_trigger
WHERE tgrelid = 'player_record'::regclass;
\end{lstlisting}

\begin{lstlisting}[language=SQL, caption={An example of trigger function on table \texttt{player\_record}}, label={lst:sample_trigger}]
CREATE OR REPLACE FUNCTION update_squad_size_transfer()
RETURNS TRIGGER AS $$
BEGIN
    -- Decrement squad size from old club
    IF OLD.club_code IS NOT NULL THEN
        UPDATE player_record
        SET squad_size = squad_size - 1
        WHERE club_code = OLD.club_code;
    END IF;

    -- Increment squad size for new club
    IF NEW.club_code IS NOT NULL THEN
        UPDATE player_record
        SET squad_size = squad_size + 1
        WHERE club_code = NEW.club_code;
    END IF;

    RETURN NEW;
END;


\end{lstlisting}

\section{LLM Prompts Examples}
\begin{tcolorbox}[title={\name {} Prompt}]
\scriptsize
\begin{lstlisting}
Database Schema:
{schema}

Instruction:
{instruction}

Generate an UPDATE SQL statement to update the player's club_name and club_code columns, and do consider the ripple effects via this update, since this update may cause other columns update. Other columns, if required, should be handled via subqueries, you will never know the content of the data table except table schema.

Think about this step by step, and you need just one SQL UPDATE query (could be with subqueries) as output.
1. First, what are the columns needed to be updated for this table schema? Come up with a UPDATE skeleton with columns need to update, no LIMIT is needed in outer skeleton.
2. Second, query each column data needed to be used for each columns update, remember to use the LIMIT clause in SUBQUERY since the subquery is used to fill the outer skeleton.
3. Combine the previous two queries.

Carefully follow these rules for SQL formatting:
- Use double quotes for all SQL identifiers (table names, column names).
- Use single quotes around all literal string values (such as player codes or club names).

Think step-by-step and return exactly one SQL UPDATE query as output. Please only return the SQL statement in a code block and do not generate anything else. 
\end{lstlisting}
\end{tcolorbox}

\begin{tcolorbox}[title={Soccer Transfer Instruction}]
\scriptsize
\begin{lstlisting}
Player '{first_name} {last_name}' (code: {player_code}) changed club from {from_club_code} to {dest_club_code}. Update his/her information.
\end{lstlisting}
\end{tcolorbox}

\begin{tcolorbox}[title={Baseline Prompt}]
\scriptsize
\begin{lstlisting}
Database Schema:
{schema}

Instruction:
{instruction}

You need just one SQL UPDATE query as output. Please only return the SQL statement in a code block and do not generate anything else. 
\end{lstlisting}
\end{tcolorbox}

\begin{tcolorbox}[title={Multi-SQL Prompt}]
\scriptsize
\begin{lstlisting}
Database Schema:
{schema}

Instruction:
{instruction}

Sample table data:
{table_sample_content}


Generate an UPDATE SQL statement to update the player's club_name and club_code columns, 
# Chain of Thought:
1. Identify which table and columns to update.
2. Determine the WHERE clause to target the correct rows.
3. Compute any needed values via subqueries if necessary (e.g., aggregations).
4. Assemble into one valid UPDATE statement.

Think step-by-step and return exactly one SQL UPDATE query as output. Please only return the SQL statement in a code block and do not generate anything else. 
\end{lstlisting}
\end{tcolorbox}

\begin{tcolorbox}[title={UCI Retail Instruction}]
\scriptsize
\begin{lstlisting}
Now a return / refund case happened, stockcode is '{StockCode}', quantity in this case is {Quantity}, and its invoicedate is at {InvoiceDate}, unitprice: {UnitPrice}, and it happend in {Country}. Update this transaction in the summary table.
\end{lstlisting}
\end{tcolorbox}

\section{LLM Generated SQL Examples}
\begin{tcolorbox}[title={ChatGPT-4o with Castle on UCI Retail}]
\scriptsize
\begin{lstlisting}
UPDATE online_retail_quarterly_summary
SET 
    "quantity" = "quantity" + (
        SELECT SUM("quantity") 
        FROM (
            SELECT -1 AS "quantity"
            WHERE "stockcode" = '84978' AND "country" = 'United Kingdom' AND "invoicedate" BETWEEN '2011-10-01' AND '2011-12-31'
            LIMIT 1
        ) AS subquery_quantity
    ),
    "y2011q4_quantity" = "y2011q4_quantity" + (
        SELECT SUM("quantity") 
        FROM (
            SELECT -1 AS "quantity"
            WHERE "stockcode" = '84978' AND "country" = 'United Kingdom' AND "invoicedate" BETWEEN '2011-10-01' AND '2011-12-31'
            LIMIT 1
        ) AS subquery_y2011q4
    )
WHERE "stockcode" = '84978' AND "country" = 'United Kingdom';
\end{lstlisting}
\end{tcolorbox}

\begin{tcolorbox}[title={ChatGPT-4o with Multi-SQL on UCI Retail}]
\scriptsize
\begin{lstlisting}
UPDATE online_retail_quarterly_summary
SET 
    quantity = quantity - 1,
    y2011q4_quantity = y2011q4_quantity - 1
WHERE 
    stockcode = '84978' AND 
    country = 'United Kingdom';
\end{lstlisting}
\end{tcolorbox}

\begin{tcolorbox}[title={ChatGPT-4o with Baseline on UCI Retail}]
\scriptsize
\begin{lstlisting}
UPDATE online_retail_quarterly_summary
SET y2011q4_quantity = y2011q4_quantity - 1
WHERE stockcode = '84978' AND country = 'United Kingdom';
\end{lstlisting}
\end{tcolorbox}

\end{document}